\documentclass[sigconf]{acmart}
\AtBeginDocument{%
  }
\usepackage{subfiles}
\usepackage{tikz}
\usepackage[table]{xcolor}
\usetikzlibrary{shapes.geometric, arrows.meta, positioning, calc,}
\usepackage[dvipsnames]{xcolor}
\usepackage{balance}
\copyrightyear{2026}
\acmYear{2026}
\setcopyright{cc}
\setcctype{by}
\acmConference[SIGIR '26]{Proceedings of the 49th International ACM SIGIR Conference on Research and Development in Information Retrieval}{July 20--24, 2026}{Melbourne, VIC, Australia}
\acmBooktitle{Proceedings of the 49th International ACM SIGIR Conference on Research and Development in Information Retrieval (SIGIR '26), July 20--24, 2026, Melbourne, VIC, Australia}
\acmDOI{10.1145/3805712.3809860}
\acmISBN{979-8-4007-2599-9/2026/07}

\begin{document}

\title{FAST-MEL: A Fast, Accurate, and Storage Efficient Solution for Multimodal Entity Linking}

\author{Thomas Derrien}
\email{thomas.derrien@irisa.fr}
\orcid{0009-0004-8942-0796}
\affiliation{%
  \institution{Univ. Rennes, INSA Rennes, CNRS, Inria, IRISA - UMR 6074}
  \city{Rennes}
  \country{France}
}
\author{Laurent Amsaleg}
\email{laurent.amsaleg@irisa.fr}
\orcid{0000-0003-0204-0930}
\affiliation{%
  \institution{Univ. Rennes, CNRS, Inria, IRISA - UMR 6074}
  \city{Rennes}
  \country{France}
}
\author{Pascale Sébillot}
\email{pascale.sebillot@irisa.fr}
\orcid{0000-0002-5429-4302}
\affiliation{%
  \institution{Univ. Rennes, INSA Rennes, CNRS, Inria, IRISA - UMR 6074}
  \city{Rennes}
  \country{France}
}

\begin{abstract}
Multimodal entity linking (MEL) is the task that consists of matching textual and visual mentions of entities in unstructured data to their corresponding entities in a knowledge base (KB). To be effective in large-scale practical settings, MEL systems must meet three objectives: high linking accuracy, computational efficiency, and storage efficiency, i.e., a compact yet efficient index of the KB. In this paper, we highlight that state-of-the-art systems fail to simultaneously satisfy these 3 requirements. To meet this three-fold objective, we propose FAST-MEL, a lightweight encoder-based MEL solution that relies on a novel and compact fixed-size vectorized representation of both the textual and visual information of each entity or mention. It matches the accuracy of the best systems but performs three orders of magnitude faster. It also consumes one order of magnitude less storage than the fastest systems.

\end{abstract}




\maketitle

\section{Introduction}
\begin{figure}[t!]
\centering
\resizebox{\linewidth}{!}{%
  \input{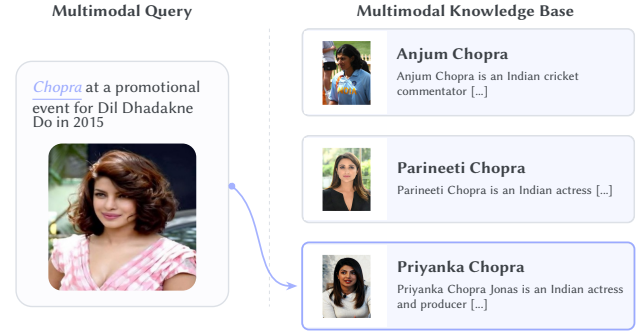}
}
\caption{Example of MEL from WikiMEL.}
\label{fig:MEL_example}
\end{figure}

Entity Linking (EL) consists of grounding entity \textit{mentions} in a document to their corresponding \textit{entities} in a knowledge base (KB). 
It can be naturally formulated as a retrieval problem over a KB, where a possibly ambiguous mention in a sentence acts as a query and the goal is to retrieve the correct entity in the KB. EL is a core component of many downstream applications such as information retrieval and question answering~\cite{xiong-etal-2019-improving,DBLP:conf/emnlp/LongprePCRD021,DBLP:conf/wsdm/MeijBO14}. Recently, Multimodal Entity Linking (MEL) \cite{DBLP:conf/acl/CarvalhoMN18} has been introduced to leverage visual context alongside text in order to reduce ambiguity and improve linking accuracy. A typical MEL setup is illustrated in Figure \ref{fig:MEL_example}.

In this setting, a multimodal query comprises a text that includes an entity mention and an associated image, while each entity in the KB is represented by a brief textual description along with an illustrative image.

Most MEL approaches~\cite{MIMIC,MELOV,OTMEL,M3EL,FissFuse,MMoE} adopt a repre\-sentation-based paradigm, in which multimodal queries and entities are projected into a shared embedding space using CLIP \cite{CLIP}, and are linked via similarity-based matching.
These encoder-based models are usually computationally lightweight and perform fast inference. However, these methods generally fall short in accuracy and require, for each entity in the KB, storing a feature vector for every textual token in its description and for every image patch in its associated image, which leads to a storage-intensive index. More recently, Large Language Models (LLMs) and Multimodal-Large-Language Models (MLLMs) have been applied to MEL. They serve several purposes: enriching the textual part of the query with synthetic context \cite{KGMEL,UNIMEL}, reformulating entity descriptions \cite{UNIMEL}, reranking a set of pre-selected candidate entities \cite{MMoE,FissFuse,KGMEL,I2CR}, or directly generating the corresponding entity name \cite{GEMEL}. These models achieve state-of-the-art accuracy by leveraging massive pretraining and reasoning capabilities, but come with higher computational costs due to large parameter counts and autoregressive inference.
They may also introduce dependencies on external APIs for closed source models. While their storage footprint can be moderate, the overall cost in terms of computing and deployment resources is not negligible.

In large-scale practical settings, MEL systems must balance three requirements:  
(1)~\textit{storage efficiency}: KBs can contain millions of entities (e.g.,  more than 6M entities of type Human in Wikidata), making compact indexing essential;
(2)~\textit{computational efficiency}: Inference must be as fast as possible;
(3)~\textit{high accuracy}: Entity linking must be as accurate as possible.

Despite its importance, this triple trade-off hasn't been studied in prior work.
In this paper, we study MEL methods from this perspective. We first highlight that existing approaches fail to satisfy all three requirements simultaneously. To meet this three-fold objective, we propose FAST-MEL\footnote{Code available at: \url{https://github.com/to2002td-cpu/FASTMEL.git}}, an encoder-based architecture built upon CLIP. Our approach relies on a novel, compact, fixed-size representation that fuses token- and patch-level information into a single 512-dimensional vector per query or entity. 
As a result, it requires one order of magnitude less storage than the fastest existing solutions (i.e., encoder-based systems). FAST-MEL also matches the accuracy of the best-performing systems (i.e., LLM/MLLM-based approaches) while running three orders of magnitude faster. By providing a lightweight, accurate and storage-efficient solution, our approach offers a practical alternative for large-scale MEL systems where deployment costs are critical.

\section{Task Definition and Analysis of Existing Systems}
The MEL task is formulated as follows. Let $q = \{q_m, q_t, q_v\}$ a multimodal query where $q_m$ is the textual mention of an entity, $q_t$ is its textual context (i.e., the surrounding words), and $q_v$ is the visual context (i.e., the image accompanying the sentence containing the mention), and a knowledge base $\mathcal{E} = \{e^i\}_{i=1}^K$ of $K$ entities. Each entity $e^i = \{e^i_n, e^i_d, e^i_v\}$ consists of a name $e^i_n$, a textual description $e^i_d$, and a visual context $e^i_v$. The goal of MEL is to link the query $q$ to the correct entity $e^* \in \mathcal{E}$.

The memory footprint of an entity in the knowledge base is defined as: $
 \textit{Entity Size (bytes)} = N_f \times d \times 2   
$, where $N_f$ is the number of features per entity, $d$ is the dimensionality of each feature, and we assume \texttt{float16} storage (2 bytes per value). Although additional compression techniques (quantization, pruning, etc.) could be used, this paper focuses on architectural choices that reduce $N_f$ and $d$.
The total KB index size is then obtained by multiplying this per-entity size by the number of entities.

In Table~\ref{tab:mel_requirements}, we compare 13 state-of-the-art MEL models through the lens of the storage–computation–accuracy trade-off. 
\begin{table*}[!ht]
\centering
\footnotesize
\caption{Comparison of 13 MEL Models under the Storage–compute–accuracy Trade-off. (\textsuperscript{†}No vector index is stored)}
\label{tab:mel_requirements}
\scriptsize
\begin{tabular}{lccccccccccccc|c}
\toprule
\textbf{Requirement} 
& MIMIC 
& OTMEL 
& MELOV 
& M3EL 
& FissFuse 
& MMoE 
& KGMEL 
& GEMEL\textsuperscript{†} 
& MMoE+DME 
& FissFuse+KAR 
& KGMEL+RR 
& I2CR\textsuperscript{†} 
& UniMEL 
& \textbf{FAST-MEL} \\
\midrule
\textbf{High accuracy }      
& $\times$ & $\times$ & $\times$ & $\times$ & $\times$ & $\times$
& $\times$ & $\times$ & $\times$ & $\times$
& $\checkmark$ & $\checkmark$ & $\checkmark$ & $\checkmark$ \\

\textbf{Comput. efficiency}         
& $\checkmark$ & $\checkmark$ & $\checkmark$ & $\checkmark$ & $\checkmark$ & $\checkmark$
& $\times$ & $\times$ & $\times$ & $\times$
& $\times$ & $\times$ & $\times$ & $\checkmark$ \\

\textbf{Storage efficiency}
& $\times$ & $\times$ & $\times$ & $\times$ & $\times$ & $\times$
& $\checkmark$ & -- & $\times$ & $\times$
& $\checkmark$ & -- & $\checkmark$ & $\checkmark$ \\
\bottomrule
\end{tabular}
\end{table*}
We propose the following classification criteria:
\textit{High accuracy} refers to models that rank within the top quartile in terms of average H@1 performance on three widely used MEL datasets (namely, WikiDiverse~\cite{WIKIDIVERSE}, RichpediaMEL~\cite{WIKIMEL_RICHPEDIAMEL}, and WikiMEL \cite{WIKIMEL_RICHPEDIAMEL}).
\textit{Storage efficiency} refers to models that do not require storing representations for all textual tokens and image patches, thereby enabling a smaller KB index size.
\textit{Computational efficiency} refers to non-generative approaches (i.e., models not based on LLMs or MLLMs), as generative frameworks typically demand substantially more GPU memory due to their larger parameter counts and incur additional computational overhead and latency from autoregressive generation. Empirical evidence supporting this claim is provided in Section~\ref{sec:exp}.

Some observations from the table follow. MIMIC~\cite{MIMIC}, MELOV~\cite{MELOV}, M3EL~\cite{M3EL}, MMoE~\cite{MMoE}, FissFuse~\cite{FissFuse}, and OTMEL~\cite{OTMEL} are computationally efficient, but sacrifice accuracy and storage efficiency. KGMEL~\cite{KGMEL}, MMoE+DME~\cite{MMoE}, KGMEL+RR~\cite{KGMEL}, GEMEL~\cite{GEMEL}, FissFuse+KAR~\cite{FissFuse}, I2CR~\cite{I2CR}, and UniMEL~\cite{UNIMEL} all depend on LLMs or MLLMs. They are therefore not computationaly efficient, although some are storage-efficient. Moreover, KGMEL, MMoE+DME, KGMEL+RR use OpenAI GPT-3.5-turbo or GPT-4o-mini and thus incur API dependence and cost.

Table~\ref{tab:mel_requirements} highlights the trade-offs in MEL system design: models that achieve high accuracy tend to use large generative models, while those emphasizing storage compactness often sacrifice performance and storage efficiency. Notably, no existing model meets all three requirements simultaneously, motivating the need for an approach that accounts for all of them.

\section{FAST-MEL: Fast, Accurate and STorage Efficient MEL}
In this section, we describe our method for satisfying all three requirements in three stages: (1) \textit{feature encoding}, which extracts token- and patch-level representations for queries and entities; (2) \textit{feature pooling}, where we introduce a novel strategy that aggregates all features to reduce storage overhead; and (3) \textit{contrastive training}, which incorporates a new hard-negative strategy to learn more discriminative representations and improve performance.

\subsection{Feature Encoding}

Following prior work \cite{MIMIC,MELOV,M3EL,OTMEL,FissFuse,MMoE}, we first extract textual and visual features for both queries and entities using pre-trained encoders. 
The textual and visual encoders are denoted by $E_t$ and $E_v$ respectively.
For text encoding, the following templates are used:
\begin{equation}
\begin{aligned}
  F^q_t &= E_t(\texttt{[CLS]}~q_m~\texttt{[EOT]}~q_t~\texttt{[EOT]}),\\
  F^e_t &= E_t(\texttt{[CLS]}~e_n~\texttt{[EOT]}~e_d~\texttt{[EOT]}).
\end{aligned}
\label{eq:feature_encoding}
\end{equation}

\noindent Here, $F^q_t$ and $F^e_t \in \mathbb{R}^{S \times d_t}$ are, respectively, the query and entity token-level embeddings of size $d_t$. $S$ denotes the length of the token sequence. Similarly, visual patch-level representations are obtained by passing the images through the visual encoder:
\begin{equation}
F^q_v = E_v(q_v), \quad F^e_v = E_v(e_v),
\label{eq:feature_enc_img}
\end{equation}
\noindent where $F^q_v$ and $F^e_v \in \mathbb{R}^{P \times d_v}$ are the patch-level $d_v$-dimensional embeddings of the query and entity images. $P$ is the number of image patches, including a \texttt{[CLS]} patch at the beginning of the sequence.

\subsection{Feature Pooling}
To leverage both token- and patch-level information while keeping representations compact, we propose a simple yet effective pooling strategy. This new approach combines all information into a single vector per query and entity, yielding a compact embedding that preserves task-relevant information. Importantly, it avoids storing feature vectors for all $S$ textual tokens and $P$ image patches (typically 40 tokens per text and 50 patches per image). Since the strategy is identical for entities and queries, we denote their textual and visual features as $F_t$ and $F_v$, respectively. First, all 
features are projected into a shared embedding space of dimension $d$ using learned linear projections:
\begin{equation}
\hat{F}_t = F_t W_t + b_t \in \mathbb{R}^{S \times d}, \quad 
\hat{F}_v = F_v W_v + b_v \in \mathbb{R}^{P \times d},
\end{equation}

\noindent where $W_t \in \mathbb{R}^{d_t \times d}$, $W_v \in \mathbb{R}^{d_v \times d}$, $b_t, b_v \in \mathbb{R}^{d}$ are learnable parameters. After projection, the textual and visual features can be described as a set $F$ of length $L = S + P$ of feature vectors:
\begin{equation}
F = \{ f_i \mid f_i \in \hat{F}_t \ \text{or} \ f_i \in \hat{F}_v \}, \quad |F| = L.
\label{eq:set}
\end{equation}
Given this set, our goal is to retain only the most useful information. To achieve this, we propose learning a weighted average of the features. 
Each vector in the set $F$ is therefore assigned a score using a two-layer MLP with a nonlinear $tanh$ activation $\phi$:
\begin{equation}
s_i = W_2^\top \phi(W_1 f_i + b_1) + b_2, \quad i = 1, \dots, L,
\end{equation}

\noindent where $W_1 \in \mathbb{R}^{h \times d}$, $W_2 \in \mathbb{R}^{h}$, $b_1 \in \mathbb{R}^{h}$, and $b_2 \in \mathbb{R}$ are learnable parameters, and $h$ is the hidden size of the MLP. The resulting scalar scores are normalized using a softmax function to obtain non-negative coefficients that sum to one. The final pooled vector $\tilde{F} \in \mathbb{R}^{d}$ is computed as the weighted sum of the features:
\begin{equation} 
\tilde{F} = \sum_{i=1}^{L} \alpha_i f_i, \quad \alpha_i = \frac{\exp(s_i)}{\sum_{j=1}^{L} \exp(s_j)}.
\label{eq:pooling}
\end{equation}
\noindent This procedure compresses both textual token-level and visual patch-level information of an entity or a query into a single fixed-dimensional vector. It also allows the model to weight each token and patch embeddings separately. At indexing time, only this pooled feature needs to be stored. 

\subsection{Contrastive Training with Hard Negatives}

FAST-MEL is trained to learn dense representations for queries and entities using a contrastive learning framework, which encourages the model to assign higher similarity scores to correct query-entity pairs while pushing apart incorrect ones. Let $\tilde{F}^q$ denotes the pooled representation of a query, and $\tilde{F}^e$ the representation of an entity. The matching score between a query $q$ and an entity $e \in \mathcal{E}$ is computed using cosine similarity between their pooled representations:
\begin{equation}
s(q,e) = \cos(\tilde{F}^q, \tilde{F}^e) = \frac{(\tilde{F}^q)^\top \tilde{F}^e}{\|\tilde{F}^q\| \, \|\tilde{F}^e\|}.
\end{equation}
Training is performed in mini-batches, where for each query, all non-matching entities in the batch serve as \emph{in-batch negatives}. To further improve discriminability, we introduce \emph{hard negative} entities for each query. These hard negatives are mined offline from the KB comparing each query to all entities using BM25~\cite{bm25}, 
a classical sparse retrieval method. This incurs minimal preprocessing 
overhead prior to training (under 30 seconds on a standard CPU for 
15k+ examples).
The query-entity contrastive loss is defined as a cross-entropy \cite{Oord2018RepresentationLW} over the batch, including both in-batch and hard negatives, with a learnable temperature $\tau$:
\begin{equation}
\mathcal{L} = - \sum_{q \in B} \log 
\frac{\exp(\tau \, s(q, e^*_q))}{\sum_{e \in (E \cup H_q) \setminus \{e^*_q\}} \exp(\tau \, s(q,e))},
\end{equation}
where $e^*_q$ denotes the correct entity for query $q$, $H_q$ is the set of hard negatives for $q$, and $E$ is the set of entities in the batch. During training, we randomly sample one hard negative from the top 10 most similar entities for each query at each step. Minimizing this loss encourages the model to bring correct query-entity pairs closer in the embedding space while pushing apart both in-batch and hard negative entities, resulting in more discriminative representations.

\section{Experiments}\label{sec:exp}
FAST-MEL is evaluated on three public MEL datasets: WikiDiverse, RichpediaMEL, and WikiMEL. For fair comparison with prior work, we use the same dataset splits and the same subsets of Wikidata for our KB. We report standard evaluation metrics, including Hits@1 (H@1) and Mean Reciprocal Rank (MRR), to assess entity linking performance. Our model is trained end-to-end with a batch size of 64 and a learning 
rate of $1 \times 10^{-5}$, selected via a 
grid search over batch sizes $\{16, 32, 64, 128\}$ and learning rates
$\{1 \times 10^{-4}, 1 \times 10^{-5}, 1 \times 10^{-6}\}$. Regarding training time, 
FAST-MEL converges in 20--30 minutes on a single A100 40GB GPU. Results for our model are reported as the average over three runs, including the standard deviation. FAST-MEL is compared against two types of baselines: (1) encoder-based methods \cite{MIMIC,MELOV,OTMEL,M3EL,FissFuse,MMoE}, which are fast but require a large KB index size, and (2) LLM/MLLM-based approaches \cite{FissFuse,MMoE,KGMEL,GEMEL,I2CR,UNIMEL}, which are slower but often achieve higher accuracy. Note that DME, KAR, and RR refer to LLM-based modules added on top of encoder-based methods to improve accuracy. 

Table \ref{tab:bigtable} presents the performance of our method and the baseline models across the three benchmarks. All baseline results are directly reported from their respective original publications.
\begin{table*}[ht!]

\footnotesize
\caption{Comparison of MEL Models on WikiDiverse, RichPedia-MEL, and WikiMEL.
(All baseline values are taken from the original publications;
`--' Value not reported; 
\textsuperscript{†} Models using GPT-3.5-turbo; 
\textsuperscript{*} Models using GPT-4o-mini)}
\begin{tabular}{llcc cccccc c}
\toprule
\multicolumn{1}{c}{} & 
\multicolumn{1}{c}{} & 
\multicolumn{1}{c}{} & 
\multicolumn{1}{c}{\textbf{Entity}} &
\multicolumn{2}{c}{\textbf{WikiDiverse}} &
\multicolumn{2}{c}{\textbf{RichPedia-MEL}} &
\multicolumn{2}{c}{\textbf{WikiMEL}} &
\multicolumn{1}{c}{} \\

\textbf{Model} & \textbf{Venue} & \textbf{\#Params} & \textbf{ Size (bytes)}$\downarrow$ & H@1$\uparrow$ & MRR$\uparrow$ & H@1$\uparrow$ & MRR$\uparrow$ & H@1$\uparrow$ & MRR$\uparrow$ & Avg H@1$\uparrow$ ($\Delta$/FAST-MEL) \\

\midrule

\rowcolor{gray!10} \multicolumn{11}{l}{\textbf{Encoder-Only (Not Storage Efficient)}} \\
MIMIC    & KDD    & 150M & 17 280 &  63.51 &73.44& 81.02 &86.95& 87.98 &91.82& 77.50 (-8.58) \\
OTMEL    & ACL    & 150M & 17 280 &  66.07 &76.57& 83.30 &88.80& 88.97 &92.32& 79.45 (-6.63) \\
MELOV    & ACL    & 150M & 17 280 &  67.32 &81.29& 84.14 &88.26& 88.91 &92.30& 80.12 (-5.96) \\
M3EL     & KDD    & 150M & 17 280 &  70.36 &81.72& 82.82 &89.04& 88.84 &92.53& 80.67 (-5.41) \\
FissFuse & ACM MM & 150M & 19 200 &  80.30 &88.11& --    &--& 84.80 &90.26& 82.55 (-3.53) \\
MMoE     & KDD    & 150M & 17 280 &  74.59 &75.43& 84.36 &88.27& 89.07 &92.59& 82.67 (-3.41) \\
\cmidrule(lr){1-11}

\rowcolor{gray!10} \multicolumn{11}{l}{\textbf{Encoder-Only (Pooled)}} \\
\textbf{FAST-MEL} &  & \textbf{150M} & \textbf{1024} 
& \textbf{84.48}$\scriptstyle \pm 0.06$ & \textbf{89.18}$\scriptstyle \pm 0.09$
& \textbf{84.30}$\scriptstyle \pm 0.73$ & \textbf{88.12}$\scriptstyle \pm 0.53$
& \textbf{89.46}$\scriptstyle \pm 0.21$ & \textbf{92.78}$\scriptstyle \pm 0.12$
& \textbf{86.08} \\

\cmidrule(lr){1-11}
\rowcolor{gray!10} \multicolumn{11}{l}{\textbf{Using LLM/MLLM (Slow Inference)}} \\
KGMEL        & SIGIR   &  \textsuperscript{*}  & 1024   &  82.12 &86.00& 76.40 &80.94& 87.29 &89.99& 81.94 (-4.14) \\
GEMEL        & COLING  & 7B                       & --     &  86.30 &--& -- &--& 82.60 &--& 84.45 (-1.63) \\
MMoE+DME     & KDD     & \textsuperscript{†}    & 17 280 &  77.57 &84.23& 85.51 &89.86& 90.77 &93.75& 84.62 (-1.46) \\
FissFuse+KAR & ACM MM  & 7B                       & 19 200 &  83.29 &89.81& -- &--& 87.89 &92.02& 85.59 (-0.49) \\
KGMEL+RR     & SIGIR   & \textsuperscript{†*}   & 1024   &  88.23 &90.84& 85.21 &88.08& 90.58 &93.04& 88.01 (+1.93) \\
I2CR         & ACM MM  & 8B                       & --     &  92.60 &--& 94.70 &--& 86.80 &--& 91.37 (+5.29) \\
UniMEL       & CIKM    & 15B                      & 8192   &  92.90 &--& 94.80 &--& 94.10 &--& 93.93 (+7.85) \\

\bottomrule
\end{tabular}
\label{tab:bigtable}

\end{table*}
FAST-MEL achieves an average H@1 score of 86.08, outperforming all encoder-based methods by a margin of at least three points. In addition, it is the most storage-efficient, with an index size close to $17\times$ smaller than encoder-based approaches, and on par with the most storage-economical solution, KGMEL. 
Compared with LLM/MLLM-based approaches, FAST-MEL surpasses the majority of them (4 out of 7). The three methods that achieve higher performance—KGMEL+RR, I2CR, and UniMEL—incur substantially greater computational costs and latency, which we discuss below.

Table \ref{tab:Comp_eff} compares the online inference time of FAST-MEL with the strongest encoder-based baseline (MMoE) and the best LLM/MLLM-based approach (UniMEL). Since all 
encoder-based baselines share the same CLIP backbone, MMoE is 
representative of their inference regime. Similarly, UniMEL is 
representative of LLM/MLLM-based approaches, which all rely on large 
generative models and autoregressive decoding, observation further 
supported by I2CR, which reports inference times similar to 
UniMEL~\cite{I2CR}.
\begin{table}[h]
\caption{Average Online Inference Time per Query on WikiMEL. (`--' indicates that a step is not applicable to the corresponding model; `Aug.' stands for augmentation)}
    \centering
    \footnotesize
    \begin{tabular}{lrrr}
    \toprule
        \textbf{Step} & \textbf{UniMEL} & \textbf{MMoE} & \textbf{FAST-MEL} \\
    \midrule

        Query aug. with image (LLaVa) (ms) & 1122.92 & -- & -- \\ 
        Query textual aug. (LLaMa) (ms) & 85.33 & -- & -- \\ 
        Embedding creation (ms) & 81.14 & 1.14 & 1.11 \\ 
        Top-K retrieval (ms) & 49.52 & -- & -- \\ 
        Final scoring (ms) & 119.74 & 4.16 & 0.15 \\
        \cmidrule(lr){1-4}
        Avg total matching time per query (ms) & 1458.65 & 5.30 & 1.26 \\
    \bottomrule
    \end{tabular}

    \label{tab:Comp_eff}
\end{table}
The experiment is conducted on 1,000 queries and 1,000 entities 
sampled from WikiMEL, using a batch size of 16 on a single A100 40GB 
GPU. We report the average time required to retrieve the most probable 
KB entity for each query. FAST-MEL achieves the lowest end-to-end matching time (1.26 ms), making it orders of magnitude faster than UniMEL (1.46 s) while also outperforming the best encoder-based model. We also evaluated the other most storage-efficient method, 
KGMEL, in the same setup (results not shown in the table), and found 
that its LLM-based generation of synthetic RDF triples for each query 
caused a substantial overhead, exceeding 4 seconds per query. This 
highlights the critical bottleneck of these generative methods compared 
to encoder-based approaches. 

\begin{figure} 
    \centering
\includegraphics[width=0.75\linewidth]{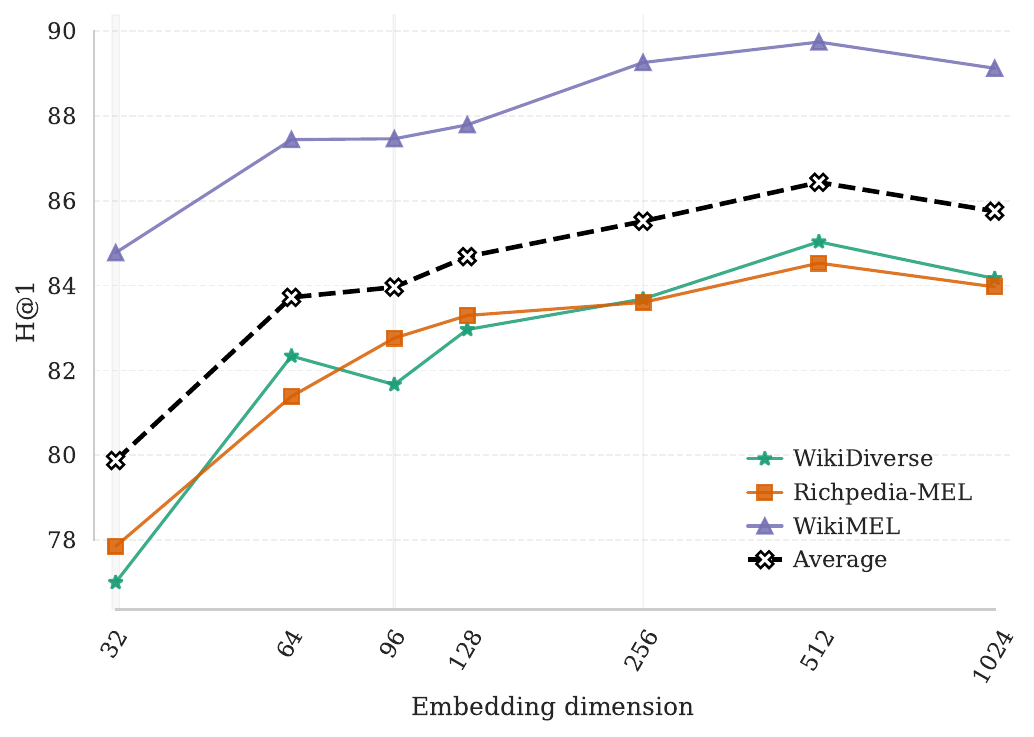}
    \caption{Impact of the Final Embedding Dimension $d$ on H@1 Performance.}
    \Description{Line graph showing the relationship between embedding dimension (x-axis, logarithmic scale from 32 to 1024) and H@1 performance (y-axis, ranging from 78 to 90 percent) across three MEL datasets. Four lines are plotted: WikiDiverse (green), Richpedia-MEL (orange), WikiMEL (blue), and their Average (black). All lines show an upward trend as dimension increases, with performance improving from 32 to 512 dimensions, then diminishing beyond 512.}
    \label{fig:dim_analysis}
\end{figure}

To further analyze our design choices, we conducted an ablation study (Table \ref{tab:ablation_study}).
\begin{table}
\caption{Ablation Study}
\label{tab:ablation_study}
\footnotesize
\centering
\begin{tabular}{lccc c}
\toprule
\textbf{Setting} & \textbf{RichP H@1} & \textbf{WikiD H@1} & \textbf{WMEL H@1} & \textbf{Avg H@1} \\
\midrule

\rowcolor{gray!10} \multicolumn{5}{l}{\textbf{Pooling Strategy}} \\
CLS-Mean & 8.71 & 20.66 & 26.29 & 18.55 \\
CLS-MLP & 9.81 & 22.62 & 27.25 & 19.89 \\
Mean Pooling & 83.73 & 81.07 & 88.12 & 84.31 \\
Max Pooling & 80.46 & 77.90 & 85.55 & 81.30 \\
\cmidrule(lr){1-5}

\rowcolor{gray!10} \multicolumn{5}{l}{\textbf{Modality}} \\
Text only & 81.05 & 85.80 & 87.29 & 84.71 \\
\cmidrule(lr){1-5}

\rowcolor{gray!10} \multicolumn{5}{l}{\textbf{Hard Negative Mining Strategy}} \\
None & 68.86 & 72.96 & 82.04 & 74.62 \\
SBERT & 82.05 & 85.10 & 89.40 & 85.52 \\
\cmidrule(lr){1-5}
\textbf{FAST-MEL }& \textbf{84.48}$\scriptstyle \pm 0.06$ 
& \textbf{84.30}$\scriptstyle \pm 0.73$ 
& \textbf{89.46}$\scriptstyle \pm 0.21$ 
& \textbf{86.08} \\
\bottomrule
\end{tabular}
\end{table}
First, we investigated different pooling strategies. \textit{CLS-Mean} averages the CLS tokens from the image and text encoders (Eq. \ref{eq:feature_encoding} and Eq.\ref{eq:feature_enc_img}). \textit{CLS-MLP} keeps only the CLS tokens in $F$ (Eq. \ref{eq:set}) and applies our learned pooling (Eq. \ref{eq:pooling}). In both cases, severe performance degradation is observed, indicating that relying solely on CLS representations fails to capture task-relevant information. In contrast, classical pooling of all token- and patch-level features using \textit{mean} or \textit{max} operators yields competitive performance. While these approaches highlight the importance of preserving token- and patch-level information, they consistently perform worse than our learned pooling method. Similarly to previous studies \cite{KGMEL,UNIMEL,FissFuse}, an image ablation was conducted to evaluate the contribution of visual features. To do so, image tokens were removed from $F$ (Eq. \ref{eq:set}). This ablation resulted in a slight performance improvement on WikiDiverse, suggesting that images possibly introduce more noise than valuable information for our model on this benchmark. Nevertheless, the full multimodal setup remains superior overall, achieving the highest average H@1 score across datasets. Regarding hard negative mining, our ablation shows that it is crucial for learning a discriminative embedding space. Removing this component reduces performance to 74.62\% Avg H@1. Trying to replace BM25 with an SBERT~\cite{SBERT} model does not improve performance but introduces a significantly higher computational cost. Finally, the sensitivity of the model to the embedding dimension $d$ was analyzed (cf.~Figure \ref{fig:dim_analysis}).
H@1 improves rapidly from $d = 32$ to $d = 256$. An optimal peak is observed at $d = 512$ (i.e., an entity size of 1024 bytes), providing a good balance between representational capacity and storage efficiency. Increasing the dimension to 1024 results in a slight decrease in performance.

\section{Conclusion}
In this article, we analyzed MEL from a new perspective: the trade-off between storage efficiency, computational efficiency, and high accuracy, which is essential for use in large-scale practical settings. Highlighting that state-of-the-art MEL systems do not meet this triple objective, we introduced FAST-MEL, a lightweight encoder-based solution that satisfies these three key requirements, relying on compact representations and hard negative mining. Extensive experiments demonstrate that FAST-MEL achieves strong performance without compromising efficiency.

\begin{acks}
This work was in part publicly funded through the French ANR (Agence
Nationale de la Recherche) under the project AGAPE with the reference
ANR-24-CE38-7253.
\end{acks}

\bibliographystyle{ACM-Reference-Format}
\balance
\bibliography{sections/refs}










\end{document}